\documentclass[conference,onecolumn]{IEEEtran}
\usepackage{graphicx}
\usepackage{comment}

\ifCLASSINFOpdf
\else
\fi
\begin{document}
\title{Consequences of Optimality}

\author{
%\IEEEauthorblockN{Michael Shell}
%\IEEEauthorblockA{School of Electrical and\\Computer Engineering\\
%Georgia Institute of Technology\\
%Atlanta, Georgia 30332--0250\\
%Email: http://www.michaelshell.org/contact.html}
%\and
\IEEEauthorblockN{Dibakar~Das}
\IEEEauthorblockA{IIIT Bangalore\\
Email: dibakard@acm.org}
%\and
%\IEEEauthorblockN{James Kirk\\ and Montgomery Scott}
%\IEEEauthorblockA{Starfleet Academy\\
%San Francisco, California 96678-2391\\
%Fax: (888) 555--1212}
}


%Telephone: (800) 555--1212, Fax: (888) 555--1212}


\maketitle

\begin{abstract}
Rationality is often related to optimal decision making. Humans are
known to be bounded rational agents. However, recent advances in
computing, and other scientific and technical fields along with large amount of data have led to a feeling that this
could result in extending the limits of bounded rationality in humans
through augmented machine intelligence. In this paper, results from a
computational model show that as more agents reach global optimality,
faster with enhanced computing, etc., solving the same problem independently, this leads to accelerated “tragedy of the commons” due to quicker resource consumption. Thus, bounded rationality could be seen as blessing in disguise (providing diversity to solutions for the same problem) from sustainability standpoint.
\end{abstract}


% \fi
\IEEEpeerreviewmaketitle

\section{Introduction}
%This demo file is intended to serve as a ``starter file''
%for IEEE conference papers produced under \LaTeX\ using
%IEEEtran.cls version 1.8 and later.
%I wish you the best of success.


%\hfill December 27, 2012
Rationality (or rather, \emph{perfect rationality}) is often relates to agents taking optimal decision always. The subject of rationality has been a much debated topic for decades. The definition of rationality varies across different subjects, such as, cognition, philosophy, psychology, economics, and mathematical and computational theories, etc. \cite{cite_scope_of_rationality}. Several theories, e.g., game theory, initially developed based on perfect rationality of players. Computationally and mathematically, perfect rationality often means maximizing agents' (e.g. humans) expected utility always reaching global optimum to a given problem \cite{cite_rationality_as_maximize_utility}.
However, perfect rationality has been countered with \emph{bounded rationality} \cite{cite_handbook_of_bounded_rationality}.
In bounded rationality, humans take decisions which are not always globally optimal  due to inherent biases and limited resources to compute. Mathematically, it is represented as a sub-optimal decision, a local optimum.

With the recent advances in science \& technology, (high performance) computing, artificial intelligence, machine learning and data science, etc., there is growing research interests on humans achieving higher levels of rationality by extending the bounds of bounded rationality
 \cite{cite_ai_will_enhance_the_bounds_bounded_rationality_marwala}.
The evolving field of \emph{computational rationality} \cite{cite_computational_rationality_review} is a step forward in this direction. Computational rationality is bringing about the convergence of subjects, such as, cognition, artificial intelligence and computational neuroscience which can lead to maximization of expected utility by agents. These advances in different domains mean that humans would be able to reach global optimal solution quicker with higher regularity and eventually always with augmentation of computational rationality.

\emph{Sustainability} has three components, namely, environment, economy and society \cite{cite_sustain_env_economy_society}. Global sustainable development has been a major thrust area for couple of decades due to concerning state of the earth's environment \cite{cite_unesco_sustainable_development}. Optimal resource utilization is critical to sustainable development.
The emerging subject of \emph{computational sustainability} deals with computational and mathematical methods to optimize usage of environmental, economic and social resources \cite{cite_computation_sustainability_intro}. Unhindered usage of resources by agents lead to exhaustion of resources which is referred to as \emph{tragedy of the commons} \cite{cite_tragedy_of_the_commons_hardin}.

This article tries to highlight the problem whether the accelerated ability to attain global optimal solutions by increasing number of agents, solving the same problem, equipped with the above mentioned advancements in different fields, lead to accelerated tragedy of the commons. A simple agent comprising a group of agents trying to solve the same problem (Knapsack problem) on its own with evolutionary method of genetic algorithms (GA), results show that as agents achieve global optimality, faster with higher computing and
data resources, this leads to accelerated tragedy of the commons.

Lets try to understand the problem with a simple example. A group of agents make pizzas independently without any communication among them from set of ingredients $\{a,b,...,z\}$. The most tasty pizza is made when the ingredients $\{a,e,i,o,u\}$ are chosen. However, lets assume that agents do not know this combination of ingredients. Hence, each agent tries to make pizzas within their local condition, knowledge  and availability of ingredients (local optima). Subsequently, a few of the agents may reach of global optimal set of $\{a,e,i,o,u\}$ and many will not. This because there are $2^{26}$ (exponential) possibilities if one bit is used to represent whether an item is chosen or not from  the set $\{a,b,...,z\}$.  In this situation, the utilization of the resources $\{a,e,i,o,u\}$ are much less stressed. However, if the agents are provided with higher capabilities, e.g., computing, algorithms and other advances in various fields mentioned above, most of the agents would reach the global optimal solution of $\{a,e,i,o,u\}$ quicker. As more and more agents would be using the resources $\{a,e,i,o,u\}$ leading to over consumption of these resources. Eventually, when all the agents using this global optimal solution would exhaust all the ingredients in the set $\{a,e,i,o,u\}$ leading to tragedy of the commons. Thus, acceleration of reaching global optimality (implying higher rationality), assisted with advances in various fields, computing, machine intelligence, etc., would lead to accelerated tragedy of the commons. Bounded rationality seems to be a blessing in disguise for sustainability due to the diversity of solutions (though sub-optimal) it inherently provides rather than same global optimal solution for all agents.

Note that this paper extends the scenario where increase in consumption of resources happen due to higher demand from lowering of cost of the resource which in turn is due to increased efficiency in production of resource \cite{cite_jevons_paradox_5}. Importantly, this paper considers the very process of multiple agents independently trying to attain global optimality for the same problem faster with higher computing capabilities can lead to \emph{accelerated consumption of resources} (independent of the cost of the
resource or production efficiency) and how diversity of solutions (local optima of agents) help avoiding the same.

The paper is organized as follows. Section \ref{section_computation_model} describes the computation model. Results obtained from the model are discussed in section \ref{section_results}. Possible ways to deal with the problem is explained in section \ref{section_way_possible_solutions}. Section \ref{section_conclusion} concludes this paper with some future extensions.
\section{Computational Model}\label{section_computation_model}
The computational model consists a set of agents trying to solve the same NP-complete problem independently. Knapsack problem is a combinatorial problem which belongs to the class of NP-complete problems \cite{cite_knapsack_6}. The objective of this problem is to select a set of items (partial or whole), each with an associated value, to be put into a sack with certain capacity such that it maximizes the total value of the items in the sack. This paper considers the whole item to be put into the sack which is referred to as 0/1-knapsack problem.

Let there be $N$ agents each trying to solve on its own the same 0/1-knapsack problem of $M$ items with weight vector $[w_1, w_2, ..., w_M]$, value vector $[v_1, v_2, ..., v_M]$ and a sack of capacity $W$. The 0/1-knapsack problem is defined as % $maximize \sum_{i=1}^M x_iv_i$ subject to $\sum_{i=1}^M x_iw_i < W$
\begin{equation}\label{eqn_knapsack_objective}
maximize \sum_{i=1}^M x_iv_i
\end{equation}
subject to
\begin{equation}\label{eqn_knapsack_constraint}
\sum_{i=1}^M x_iw_i < W
\end{equation}
$x_i \in \{0,1\}$. The solution essentially boils down to selecting a vector $X^{(opt)}$ consisting of 0s and 1s from $2^M$ possibilities which maximizes $\sum_{i=1}^M x_iv_i$ satisfying constraint (\ref{eqn_knapsack_constraint}).

Knapsack is a well-studied problem which can be solved by dynamic
programming (DP) \cite{cite_knapsack_6}. Also, approaches such as GA to solve the problem have also been proposed \cite{cite_ga_knapsack_7}. Let’s assume that agents do not use (or do not have the capability to apply) DP and apply trial-and-error method to solve the problem. This trial-and-error method is modeled using a GA due to the reason that it helps emulate an evolutionary process of an agent progressively achieving improved optimality over time (generations in GA).

Each agent $j$ starts its GA with an initial random population (vectors of 1s and 0s of length M) of size $N_p^{(j)}$ as starting values of solution vector $X^{(j)}$. $N_s^{(j)}$ vectors from $N_p^{(j)}$  which satisfy the constraints are taken as possible solution candidates. Among $N_s^{(j)}$ there is a possible better local optima or the global optimum for agent $j$. If size of $N_s^{(j)} < 2$ then a new initial population $N_p^{(j)}$ is generated. Any two vectors from $N_s^{(j)}$, say, $X^{(j)(m)}$ and $X^{(j)(n)}$  are selected at random by agent $j$. For the crossover, upper half of $X^{(j)(m)}$ and lower half of $X^{(j)(n)}$ are concatenated and a new vector $X^{(j)(p)}$ is constructed. Some bits of $X^{(j)(p)}$ are
toggled in the mutation step. The steps of random selection, crossover and mutation are repeated a number of times to form a new population $N_p^{(j)}$ to be tried in the next generation. This entire process is repeated over generations $N_g^{(j)}$ by each agent $j$ till it has a satisfactory local optimum solution or reaches global optimum $X^{(opt)}$. In each generation
$k$, an agent will either reach the global optimum or have a local optimum satisfying the constraints. Over multiple generations, each agent improves its current solution with a better local optimum or achieves global optimum.

Let’s assume that for each item $i$ with weight $w_i$, $i = 1, 2,.., M$, there is $\alpha_i r_i$ amount of resource consumed when item $i$ is in the solution (local or global optimum), where $r_i$ is a resource and $\alpha_i$ is a scaling factor. Then, the cumulative resources consumed over all the generations is %$\sum_{i=1}^M\sum_{j=1}^N\sum_{l=1}^{N_g^{(j)}}x_i^{(j)(k)}\alpha_ir_i$,
\begin{equation}
\sum_{i=1}^M\sum_{j=1}^N\sum_{k=1}^{N_g^{(j)}}x_i^{(j)(k)}\alpha_ir_i
\end{equation}
where $x_i^{(j)(k)} \in \{0,1\}$.
If the cumulative resource of $r_i$ consumed in any generation is greater than total availability of $r_i$ then all the agents which uses that particular resource (as part of local or global optimum) would continue to starve thereafter, leading to tragedy of the commons. The faster the agents achieve their global optima due to increased computational abilities would lead to accelerated tragedy of commons.
%\IEEEpubidadjcol

\section{Results}\label{section_results}
This section presents the simulation results using the above computation model. Parameters used in the simulation are provided in Table \ref{table_simulation_parameters}. Simulation is performed using \verb|python| language.

Three scenarios of resource consumption are considered as follows.
\begin{enumerate}
\item When all agents try to achieve global optimum.
\item When most agents are satisfied with local optimum.
\item When some agents have the capability to achieve the global optimum faster due to higher computing capabilities.
\end{enumerate}
\begin{table}[ht]\label{table_simulation_parameters}
  \caption{Simulation Parameters}
  \centering
  \begin{tabular}{|p{2cm}|p{2cm}|p{3cm}|}
  \hline
  Parameter & Value & Description\\  [0.5ex]
  \hline
  $N$ & 25 & Number of agents\\
  \hline
  $M$ & 10 & Number of items\\
  \hline
  $w_1, w_2,..., w_{10}$ & 996, 771, 543, 593, 621, 473, 595, 388, 935, 874 & Weights of items, uniform randomly generated\\
  \hline
  $v_1, v_2,..., v_{10}$ & 54.04769411, 39.33601431, 14.83657681, 43.52375770, 66.31920392, 26.17907976, 27.14489409, 58.72956010, 25.50253249, 49.04678721 & Value of items, uniform randomly generated\\
  \hline
  $W$  & $0.5 \times \sum_{i=1}^{10}w_i$ & Capacity of sack\\
  \hline
  $X^{(opt)}$ & 1,1,0,1,1,0,0,1,0,0 & Optimal solution using dynamic programming for validation only\\
  \hline
  $X^{(opt)}$ & 1,1,0,1,1,0,0,1,0,0 & Optimal solution using dynamic programming for validation only\\
  \hline
  $N_p^{(j)}, j = 1,2,..,25$  & 45 & Initial population size. Assumed same for all agents\\
  \hline
  $N_s^{(j)}, j = 1,2,.., 25$  & 45 & Probable solution set size. Assume same for all agents\\
  \hline
  $N_g^{(j)}, j = 1,2,.., 25$  & 2000 & Number of generations each agent tries to achieve better or optimal solution. Assumed same for all agents\\
  \hline
  $\alpha_i, i=1,2,..,10$ & 1 & Assumed same for all agents\\
  \hline
  $r_i, i = 1,2,..,10$ & $w_i$ & Assuming that each item consumes resources equal to its weight, $w_i = r_i$\\
  \hline
  \end{tabular}
  \label{table_model_parameters}
\end{table}

\begin{figure}[ht]
\centering
\includegraphics[width=\columnwidth]{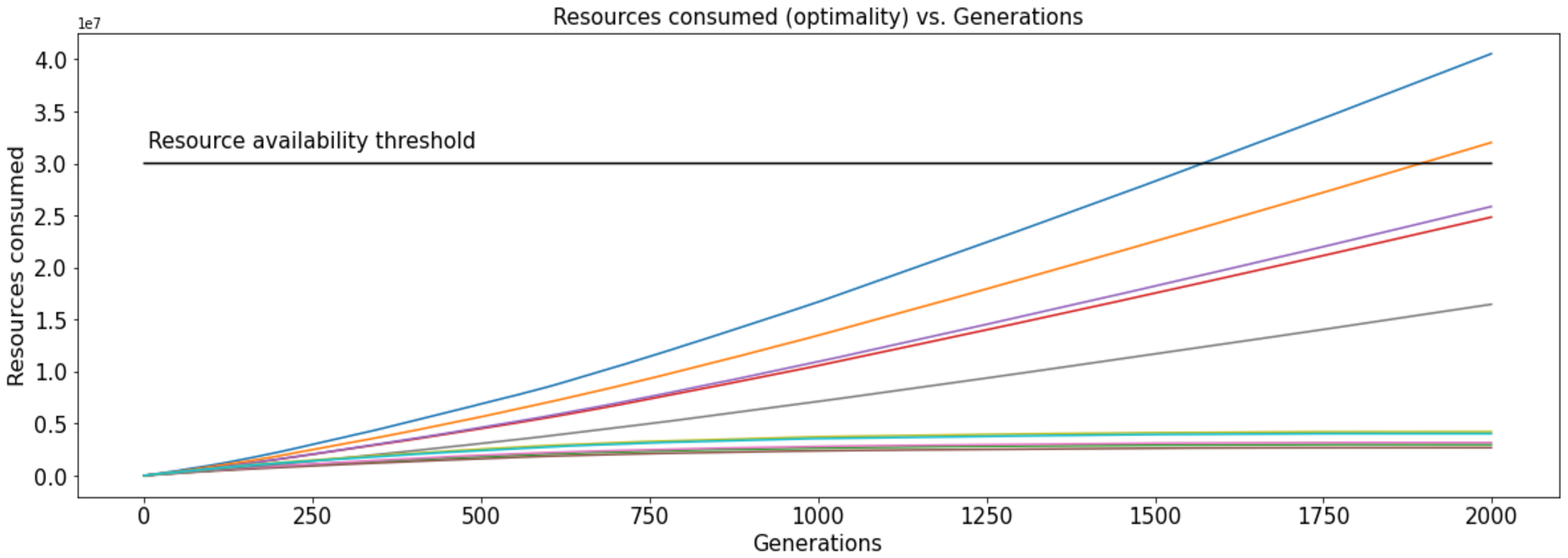}
\caption{Resource consumption (first scenario)}
\label{fig_fig1_inkscape}
\end{figure}

\begin{figure}[ht]
\centering
\includegraphics[width=\columnwidth]{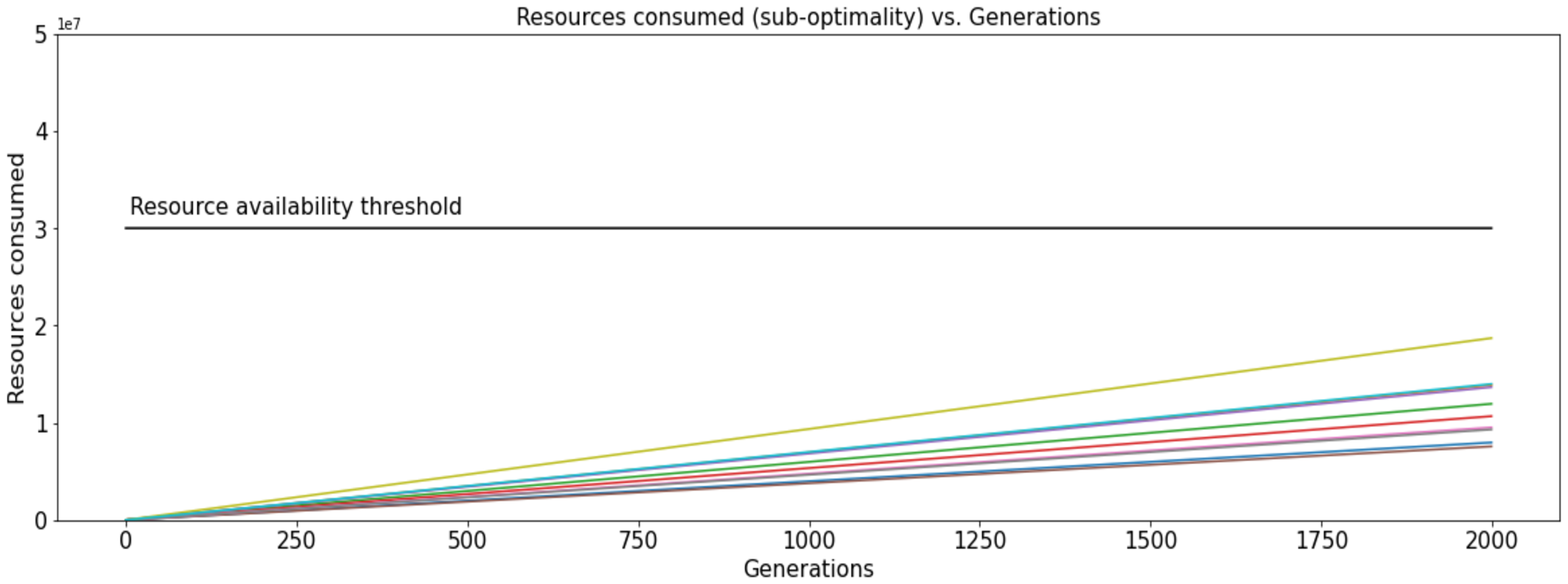}
\caption{Resource consumption (second scenario)}
\label{fig_fig2_inkscape}
\end{figure}

\begin{figure}[ht]
\centering
\includegraphics[width=\columnwidth]{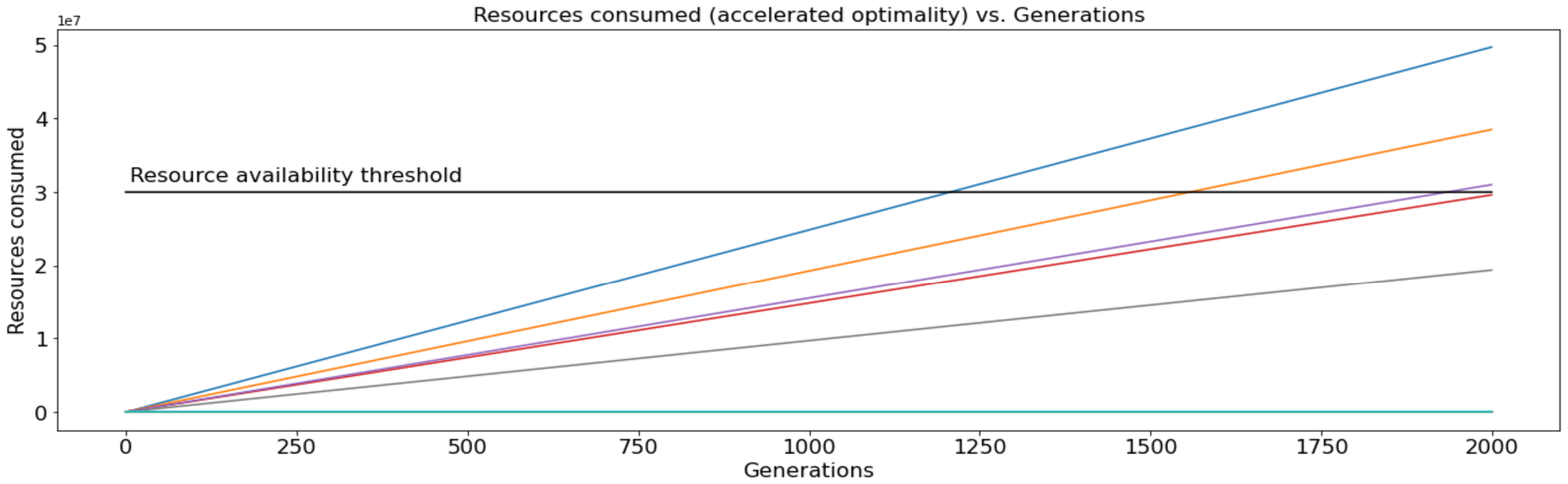}
\caption{Resource consumption (third scenario)}
\label{fig_fig3_inkscape}
\end{figure}

\begin{figure}[ht]
\centering
\includegraphics[width=\columnwidth]{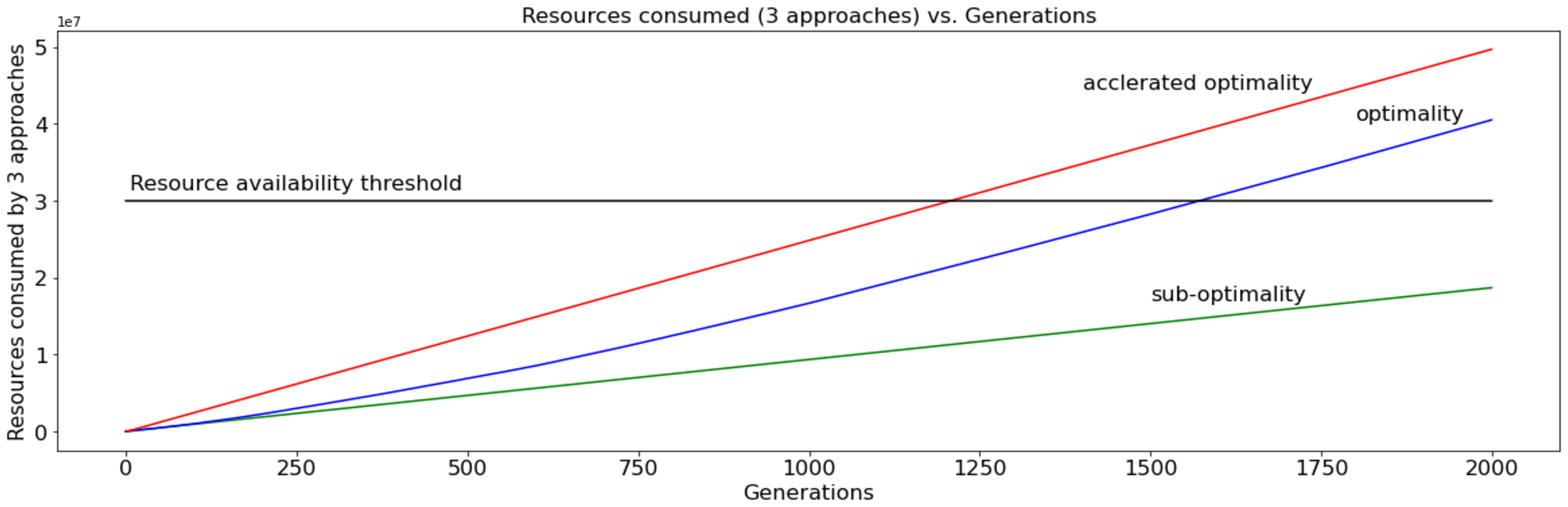}
\caption{Highest resource consumed (all 3 scenarios)}
\label{fig_fig4_inkscape}
\end{figure}

Resource consumption for the first scenario is shown in Fig. \ref{fig_fig1_inkscape}. The cumulative resources consumed are shown along \emph{y}-axis and generations are along \emph{x}-axis for all the
Figs. \ref{fig_fig1_inkscape}-\ref{fig_fig4_inkscape}. It can be observed that the highest consumed resource exceeds the resource availability threshold at around generation 1600. In this case all agents achieve global
optimum.

Resource consumption for the second scenario is shown in Fig. \ref{fig_fig2_inkscape}. Here, most agents are satisfied with the local minima. It can be observed that for the same number of generations, cumulative resources consumed by the agents is much below the resource availability threshold. In this case, none of the agents achieve global optimum.

Resource consumption for the third scenario is shown in Fig. \ref{fig_fig3_inkscape}. The axes remain the same. With some random agents achieve global optimum faster (due to their higher computing capabilities), the highest consumed resource exceeds the resource availability threshold at around generation 1200, which is much earlier than in Fig. \ref{fig_fig1_inkscape} of generation 1600. In this case also all agents achieve global optimum much earlier than Fig. \ref{fig_fig1_inkscape}.

Fig. \ref{fig_fig4_inkscape} shows the cumulative highest consumed resource extracted from Figs. \ref{fig_fig1_inkscape}-\ref{fig_fig3_inkscape} for ease of inference. Clearly, as agents strive for global optimum solving the same problem, and quicker they try to achieve the same (due to higher computing capabilities), lead to faster resource consumption, causing accelerated tragedy of the commons.
\section{Possible solutions to accelerated tragedy of the commons}\label{section_way_possible_solutions}
Given that advances in various fields mentioned above is a continuous and inevitable process, it is necessary to find ways to deal with the situation of reaching accelerated optimality (implying higher rationality) and consequential faster resource consumption. One of the common ways to deal with tragedy of commons is regulation of resource usages. Another way is to patronize local solutions with basic research to exploit local resources or suggest substitutes for scarce ones. Disruptive innovation is another possible way to bring about new techniques to replace existing ones. Interpreting the global optimal solution in local context and customizing them to suit the needs is of paramount importance from sustainable viewpoint. Achieving a local optimal solution closest to global optimum and maintaining diversity would be a  major challenge to deal with in future.
\section{Conclusion and future work}\label{section_conclusion}
Rationality has been a very intriguing topic for a long time. Perfect rationality has been contended by bounded rationality where agents have certain limitations to take the global optimal solution always. However, advances in computing and other scientific and technological fields have the possibility of extending the limits of bounded rationality eventually leading to perfect rationality where agents always achieve global optimum  with augmented machine intelligence (e.g., computation rationality).
Using an agent based computational model where agents try to solve 0/1-knapsack problem on their own applying trial-and-error approach, results show agents achieving global optimum for the same problem quicker with enhanced computing, etc., can lead to accelerated tragedy of the commons.
Thus, diversity of solutions for the same problem with local optima (due to bounded rationality) are more sustainable than one global solution from (computational) sustainability standpoint. However, there could be externalities of agents using local sub-optimal solutions which will be considered in future work. Some of the approaches suggested above to deal with this problem of accelerated resource consumption need to be investigated to deal these emerging developments.

\bibliographystyle{IEEEtran}
\bibliography{IEEEabrv,sample-base}

\end{document}